\documentclass[twocolumn,letterpaper,aps,prd,longbibliography,superscriptaddress,showpacs,floatfix]{revtex4-1}

\usepackage{graphicx}	% Include figure filed
\usepackage{amsmath}
\usepackage{longtable}
\usepackage{xspace}	% Include xspace

\begin{document}

\title{Measurement of parity-violating spin asymmetries in W$^{\pm}$ 
production at midrapidity in longitudinally polarized $p$$+$$p$ collisions}

\newcommand{\abilene}{Abilene Christian University, Abilene, Texas 79699, USA}
\newcommand{\augie}{Department of Physics, Augustana University, Sioux Falls, South Dakota 57197, USA}
\newcommand{\banaras}{Department of Physics, Banaras Hindu University, Varanasi 221005, India}
\newcommand{\barc}{Bhabha Atomic Research Centre, Bombay 400 085, India}
\newcommand{\baruch}{Baruch College, City University of New York, New York, New York, 10010 USA}
\newcommand{\bnlcoll}{Collider-Accelerator Department, Brookhaven National Laboratory, Upton, New York 11973-5000, USA}
\newcommand{\bnlphys}{Physics Department, Brookhaven National Laboratory, Upton, New York 11973-5000, USA}
\newcommand{\caucr}{University of California-Riverside, Riverside, California 92521, USA}
\newcommand{\charlesczech}{Charles University, Ovocn\'{y} trh 5, Praha 1, 116 36, Prague, Czech Republic}
\newcommand{\chonbuk}{Chonbuk National University, Jeonju, 561-756, Korea}
\newcommand{\ciae}{Science and Technology on Nuclear Data Laboratory, China Institute of Atomic Energy, Beijing 102413, People's Republic of~China}
\newcommand{\cns}{Center for Nuclear Study, Graduate School of Science, University of Tokyo, 7-3-1 Hongo, Bunkyo, Tokyo 113-0033, Japan}
\newcommand{\colorado}{University of Colorado, Boulder, Colorado 80309, USA}
\newcommand{\columbia}{Columbia University, New York, New York 10027 and Nevis Laboratories, Irvington, New York 10533, USA}
\newcommand{\czechtech}{Czech Technical University, Zikova 4, 166 36 Prague 6, Czech Republic}
\newcommand{\debrecen}{Debrecen University, H-4010 Debrecen, Egyetem t{\'e}r 1, Hungary}
\newcommand{\elte}{ELTE, E{\"o}tv{\"o}s Lor{\'a}nd University, H-1117 Budapest, P{\'a}zm{\'a}ny P.~s.~1/A, Hungary}
\newcommand{\ewha}{Ewha Womans University, Seoul 120-750, Korea}
\newcommand{\fsu}{Florida State University, Tallahassee, Florida 32306, USA}
\newcommand{\gsu}{Georgia State University, Atlanta, Georgia 30303, USA}
\newcommand{\hanyang}{Hanyang University, Seoul 133-792, Korea}
\newcommand{\hiroshima}{Hiroshima University, Kagamiyama, Higashi-Hiroshima 739-8526, Japan}
\newcommand{\howard}{Department of Physics and Astronomy, Howard University, Washington, DC 20059, USA}
\newcommand{\ihepprot}{IHEP Protvino, State Research Center of Russian Federation, Institute for High Energy Physics, Protvino, 142281, Russia}
\newcommand{\illuiuc}{University of Illinois at Urbana-Champaign, Urbana, Illinois 61801, USA}
\newcommand{\inrras}{Institute for Nuclear Research of the Russian Academy of Sciences, prospekt 60-letiya Oktyabrya 7a, Moscow 117312, Russia}
\newcommand{\instpasczech}{Institute of Physics, Academy of Sciences of the Czech Republic, Na Slovance 2, 182 21 Prague 8, Czech Republic}
\newcommand{\isu}{Iowa State University, Ames, Iowa 50011, USA}
\newcommand{\jaea}{Advanced Science Research Center, Japan Atomic Energy Agency, 2-4 Shirakata Shirane, Tokai-mura, Naka-gun, Ibaraki-ken 319-1195, Japan}
\newcommand{\jyvaskyla}{Helsinki Institute of Physics and University of Jyv{\"a}skyl{\"a}, P.O.Box 35, FI-40014 Jyv{\"a}skyl{\"a}, Finland}
\newcommand{\karoly}{K\'aroly R\'oberts University College, H-3200 Gy\"ngy\"os, M\'atrai \'ut 36, Hungary}
\newcommand{\kek}{KEK, High Energy Accelerator Research Organization, Tsukuba, Ibaraki 305-0801, Japan}
\newcommand{\korea}{Korea University, Seoul, 136-701, Korea}
\newcommand{\kurchatov}{National Research Center ``Kurchatov Institute", Moscow, 123098 Russia}
\newcommand{\kyoto}{Kyoto University, Kyoto 606-8502, Japan}
\newcommand{\labllr}{Laboratoire Leprince-Ringuet, Ecole Polytechnique, CNRS-IN2P3, Route de Saclay, F-91128, Palaiseau, France}
\newcommand{\lahorelums}{Physics Department, Lahore University of Management Sciences, Lahore 54792, Pakistan}
\newcommand{\lawllnl}{Lawrence Livermore National Laboratory, Livermore, California 94550, USA}
\newcommand{\losalamos}{Los Alamos National Laboratory, Los Alamos, New Mexico 87545, USA}
\newcommand{\lpc}{LPC, Universit{\'e} Blaise Pascal, CNRS-IN2P3, Clermont-Fd, 63177 Aubiere Cedex, France}
\newcommand{\lund}{Department of Physics, Lund University, Box 118, SE-221 00 Lund, Sweden}
\newcommand{\maryland}{University of Maryland, College Park, Maryland 20742, USA}
\newcommand{\mass}{Department of Physics, University of Massachusetts, Amherst, Massachusetts 01003-9337, USA}
\newcommand{\michigan}{Department of Physics, University of Michigan, Ann Arbor, Michigan 48109-1040, USA}
\newcommand{\muhlenberg}{Muhlenberg College, Allentown, Pennsylvania 18104-5586, USA}
\newcommand{\myongji}{Myongji University, Yongin, Kyonggido 449-728, Korea}
\newcommand{\nagasaki}{Nagasaki Institute of Applied Science, Nagasaki-shi, Nagasaki 851-0193, Japan}
\newcommand{\nara}{Nara Women's University, Kita-uoya Nishi-machi Nara 630-8506, Japan}
\newcommand{\natmephi}{National Research Nuclear University, MEPhI, Moscow Engineering Physics Institute, Moscow, 115409, Russia}
\newcommand{\newmex}{University of New Mexico, Albuquerque, New Mexico 87131, USA}
\newcommand{\nmsu}{New Mexico State University, Las Cruces, New Mexico 88003, USA}
\newcommand{\ohio}{Department of Physics and Astronomy, Ohio University, Athens, Ohio 45701, USA}
\newcommand{\ornl}{Oak Ridge National Laboratory, Oak Ridge, Tennessee 37831, USA}
\newcommand{\orsay}{IPN-Orsay, Univ.~Paris-Sud, CNRS/IN2P3, Universit\'e Paris-Saclay, BP1, F-91406, Orsay, France}
\newcommand{\peking}{Peking University, Beijing 100871, People's Republic of~China}
\newcommand{\pnpi}{PNPI, Petersburg Nuclear Physics Institute, Gatchina, Leningrad region, 188300, Russia}
\newcommand{\riken}{RIKEN Nishina Center for Accelerator-Based Science, Wako, Saitama 351-0198, Japan}
\newcommand{\rikjrbrc}{RIKEN BNL Research Center, Brookhaven National Laboratory, Upton, New York 11973-5000, USA}
\newcommand{\rikkyo}{Physics Department, Rikkyo University, 3-34-1 Nishi-Ikebukuro, Toshima, Tokyo 171-8501, Japan}
\newcommand{\saispbstu}{Saint Petersburg State Polytechnic University, St.~Petersburg, 195251 Russia}
\newcommand{\saopaulo}{Universidade de S{\~a}o Paulo, Instituto de F\'{\i}sica, Caixa Postal 66318, S{\~a}o Paulo CEP05315-970, Brazil}
\newcommand{\seoulnat}{Department of Physics and Astronomy, Seoul National University, Seoul 151-742, Korea}
\newcommand{\stonybrkc}{Chemistry Department, Stony Brook University, SUNY, Stony Brook, New York 11794-3400, USA}
\newcommand{\stonycrkp}{Department of Physics and Astronomy, Stony Brook University, SUNY, Stony Brook, New York 11794-3800, USA}
\newcommand{\tenn}{University of Tennessee, Knoxville, Tennessee 37996, USA}
\newcommand{\titech}{Department of Physics, Tokyo Institute of Technology, Oh-okayama, Meguro, Tokyo 152-8551, Japan}
\newcommand{\tsukuba}{Center for Integrated Research in Fundamental Science and Engineering, University of Tsukuba, Tsukuba, Ibaraki 305, Japan}
\newcommand{\vandy}{Vanderbilt University, Nashville, Tennessee 37235, USA}
\newcommand{\weizmann}{Weizmann Institute, Rehovot 76100, Israel}
\newcommand{\wigner}{Institute for Particle and Nuclear Physics, Wigner Research Centre for Physics, Hungarian Academy of Sciences (Wigner RCP, RMKI) H-1525 Budapest 114, POBox 49, Budapest, Hungary}
\newcommand{\yonsei}{Yonsei University, IPAP, Seoul 120-749, Korea}
\newcommand{\zagreb}{University of Zagreb, Faculty of Science, Department of Physics, Bijeni\v{c}ka 32, HR-10002 Zagreb, Croatia}
\affiliation{\abilene}
\affiliation{\augie}
\affiliation{\banaras}
\affiliation{\barc}
\affiliation{\baruch}
\affiliation{\bnlcoll}
\affiliation{\bnlphys}
\affiliation{\caucr}
\affiliation{\charlesczech}
\affiliation{\chonbuk}
\affiliation{\ciae}
\affiliation{\cns}
\affiliation{\colorado}
\affiliation{\columbia}
\affiliation{\czechtech}
\affiliation{\debrecen}
\affiliation{\elte}
\affiliation{\ewha}
\affiliation{\fsu}
\affiliation{\gsu}
\affiliation{\hanyang}
\affiliation{\hiroshima}
\affiliation{\howard}
\affiliation{\ihepprot}
\affiliation{\illuiuc}
\affiliation{\inrras}
\affiliation{\instpasczech}
\affiliation{\isu}
\affiliation{\jaea}
\affiliation{\jyvaskyla}
\affiliation{\karoly}
\affiliation{\kek}
\affiliation{\korea}
\affiliation{\kurchatov}
\affiliation{\kyoto}
\affiliation{\labllr}
\affiliation{\lahorelums}
\affiliation{\lawllnl}
\affiliation{\losalamos}
\affiliation{\lpc}
\affiliation{\lund}
\affiliation{\maryland}
\affiliation{\mass}
\affiliation{\michigan}
\affiliation{\muhlenberg}
\affiliation{\myongji}
\affiliation{\nagasaki}
\affiliation{\nara}
\affiliation{\natmephi}
\affiliation{\newmex}
\affiliation{\nmsu}
\affiliation{\ohio}
\affiliation{\ornl}
\affiliation{\orsay}
\affiliation{\peking}
\affiliation{\pnpi}
\affiliation{\riken}
\affiliation{\rikjrbrc}
\affiliation{\rikkyo}
\affiliation{\saispbstu}
\affiliation{\saopaulo}
\affiliation{\seoulnat}
\affiliation{\stonybrkc}
\affiliation{\stonycrkp}
\affiliation{\tenn}
\affiliation{\titech}
\affiliation{\tsukuba}
\affiliation{\vandy}
\affiliation{\weizmann}
\affiliation{\wigner}
\affiliation{\yonsei}
\affiliation{\zagreb}
\author{A.~Adare} \affiliation{\colorado} 
\author{C.~Aidala} \affiliation{\losalamos} \affiliation{\michigan} 
\author{N.N.~Ajitanand} \affiliation{\stonybrkc} 
\author{Y.~Akiba} \affiliation{\riken} \affiliation{\rikjrbrc} 
\author{R.~Akimoto} \affiliation{\cns} 
\author{J.~Alexander} \affiliation{\stonybrkc} 
\author{M.~Alfred} \affiliation{\howard} 
\author{K.~Aoki} \affiliation{\kek} \affiliation{\riken} 
\author{N.~Apadula} \affiliation{\isu} \affiliation{\stonycrkp} 
\author{Y.~Aramaki} \affiliation{\cns} \affiliation{\riken} 
% ...
\collaboration{PHENIX Collaboration} \noaffiliation

\date{\today}

%------------------------------------------------------------------------------|
\begin{abstract}

We present midrapidity measurements from the PHENIX experiment of large parity-violating single spin asymmetries of high transverse momentum electrons and positrons from $W^\pm/Z$ decays, produced in longitudinally polarized $p$$+$$p$ collisions at center of mass energies of $\sqrt{s}$=500 and 510~GeV.
These asymmetries allow direct access to the 
anti-quark polarized parton distribution functions due to the 
parity-violating nature of the $W$-boson coupling to quarks and 
anti-quarks. The results presented are based on data collected in 2011, 
2012, and 2013 with an integrated luminosity of 240 pb$^{-1}$, which 
exceeds previous PHENIX published results by a factor of more than 27. 
These high $Q^2$ data probe the parton structure of the proton at $W$ mass scale and provide an important addition to our understanding of the anti-quark parton helicity distribution functions at an intermediate Bjorken $x$ value of roughly $M_W/\sqrt{s}=0.16$.

\end{abstract}

	
\maketitle

%\textbf{*** page break for PRL word count ***}  \clearpage

%%%%%%%%%%%%%%%%%%%%%%%%%%%%%%%%%%%%%%%%%%%%%%%%%%%%%%%%%%%%%%%%%%%%%%%%%%%%%%%% Introduction

The determination of the contributions of partons to the spin of the 
proton has inspired significant theoretical and experimental effort for 
several 
decades~\cite{Ashman:1987hv,Jaffe:1989jz,Adeva:1998vv,Anthony:1999rm,Alekseev:2007vi,Airapetian:2004zf,Adare:2014hsq,Adamczyk:2014ozi,Adare:2010xa,STARW2009,Adamczyk:2014xyw,deFlorian:2009vb,Leader:2010rb}. 
The quark contribution to the nucleon spin has been deduced through 
measurements in polarized inclusive deep-inelastic scattering (DIS) 
and semi-inclusive deep-inelastic scattering (SIDIS) experiments~\cite{deFlorian:2009vb,Leader:2010rb,Nocera:2014gqa,Airapetian:2004zf,Alekseev:2010hc}.
Although the overall quark contribution ($\Delta \Sigma=\Delta q + \Delta 
\bar{q}$) has been well-determined through DIS experiments (in the range 
$10^{-3}<x<1$), the contributions from sea quarks separated by flavor 
(determined through SIDIS experiments) are comparatively poorly known. 
Data from HERMES and COMPASS~\cite{Airapetian:2004zf,Alekseev:2010ub} 
provide constraints on the contribution from the sea quarks, however, 
uncertainties in fragmentation functions and the low energy scales of 
fixed target experiments limit the accuracy with which these measurements 
can quantitatively determine the sea quark 
contribution~\cite{Leader:2006xc}. As such, an independent measurement 
using a different technique~\cite{Bourrely:1993dd} to determine the 
contribution from different flavors of sea quarks is desirable.

The use of $W$-boson production provides just such a solution. Parity is 
maximally-violated in the $W$ couplings to quarks and leptons, so $W^\pm$ 
production in $p$$+$$p$ collisions proceeds only by coupling to 
left-handed quarks and right-handed anti-quarks ($u_L\bar{d}_R \rightarrow 
W^+$ and $d_L\bar{u}_R \rightarrow W^-$). By measuring decay leptons in 
the final state, the flavor and helicity state of the colliding quarks can 
be 
determined~\cite{Bourrely:1993dd,Bunce:2000uv,Nadolsky:2003ga,deFlorian:2010aa}. 
Asymmetries measured in $W^{\pm}$ by reversing the helicity of a colliding proton 
are sensitive to the individual quark/anti-quark helicity parton distribution 
functions (PDFs) ($\Delta u$, $\Delta d$, $\Delta \bar{u}$ and $\Delta \bar{d}$).
Moreover, the energy scale for these 
events, of the order of the $W$-boson mass, allows for small and precisely 
calculable higher-order corrections.

We present results for the parity-violating single-spin asymmetry $A_L$ for $p+p \rightarrow W^\pm / Z + X \rightarrow e^\pm + X'$ at midrapidity from 2011--2013 PHENIX data at the Relativistic Heavy Ion Collider (RHIC). These results relate to an intermediate Bjorken $x$ value of roughly $M_W/\sqrt{s}=0.16$.
Initial measurements at RHIC in 2009 accumulated 8.6 
pb$^{-1}$ by PHENIX~\cite{Adare:2010xa} and 12 pb$^{-1}$ 
by STAR~\cite{STARW2009,Adamczyk:2014xyw}.
Here, the total integrated luminosity is 240 
pb$^{-1}$ at $\sqrt{s}=500$ GeV in 2011, and at $510$ GeV in 2012 and 
2013~\cite{schoefer12}.  Proton-beam polarizations were also considerably 
improved from $\sim$0.39 in 2009 to 0.50--0.56 in 2011-2013.

The measurements are performed with the two PHENIX central arm 
spectrometers.  Each arm covers $\left| \Delta\phi\right|=\pi/2$ in 
azimuth and $\left|\eta\right|<0.35$ in pseudorapidity. A comprehensive 
description of the PHENIX detector at RHIC can be found in~\cite{Adcox:2003zm}. 
The major detector subsystems used for this analysis are the 
electromagnetic calorimeter (EMCal) and the drift chamber/pad chamber 
tracking system. Two beam-beam counters located at  $\pm$ 144 cm from the 
collision point along the beam line and covering 
$3.1<\left|\eta\right|<3.9$ were used to define the minimum bias trigger 
and to measure the relative luminosity between different colliding bunch 
pairs.

The data were collected with an EMCal-based trigger~\cite{NIM_emc} with nominal energy threshold of 5.6 GeV, which was fully efficient for $e^\pm$ with transverse momentum $p_{T}^{e}>$10 GeV/$c$.
The $p_{T}^{e}$ was determined from the energy deposited in the EMCal with 
energy resolution $\sigma_{E}/E=8.1\%/\sqrt{E(GeV)} \oplus 4\%$. The 
energy resolution was determined from the $p_T$-dependence of the widths 
of reconstructed $\pi^0$ and $\eta$ meson mass peaks. The same 
$\pi^0$ and $\eta$ meson mass peaks were used in the energy calibration of 
the EMCal, and were continuously monitored. 
Similar to our previous analysis~\cite{Adare:2010xa} and test beam data results~\cite{NIM_emc}, the EMCal energy scale was confirmed to within 2.5\%, for the energy range analyzed with this data. A loose time-of-flight cut was applied in the analysis to remove noncollision background.

The tracking system was used for collision vertex reconstruction, track 
charge sign determination, and background suppression. The main tracking 
detector, the drift chamber (DC), spanning the radial distance 2.02--2.46~m 
from the beam line, measured the charged track bending in the axial 
magnetic field of the PHENIX central magnet, with a field integral of 1.15 
Tm. The $z$-coordinate for the tracks was obtained from the pad chambers 
situated behind the DC. Reconstructed tracks were matched with high energy 
clusters in the EMCal within a cone angle of $0.02$, retaining $>$99\% of 
real $e^{\pm}$ tracks, as determined from simulations. 
The coordinate information from both the 
calorimeter and the tracking system was used to determine the $z$-vertex 
of the event, and only events with $\left| z \right| < 30$ cm were used in 
the analysis.

The charge sign of a track was determined from the bending angle 
$\alpha_{\rm DC}$, which is inversely proportional to the track transverse 
momentum ($\alpha_{\rm DC} {\rm (mrad)}=92/p_T{\rm ~GeV}/c$). A region 
corresponding to $|\alpha_{\rm DC}|<1$ mrad was removed in order to 
minimize the possibility of charge misidentification. This led to $<$3\% 
loss of $e^{\pm}$ from $W$-boson decays. To further eliminate the charge 
sign ambiguity in the DC track reconstruction, the regions in the vicinity 
of anode wires were removed from the analysis, reducing the DC acceptance 
by $\sim$15\%. The remaining opposite charge contribution to the $W^{-}$ 
($W^{+}$) signal was 2\% (0.4\%), as determined using the DC resolution of 
1.4~mrad and $\alpha_{\rm DC}$ convoluted over the $W$ decay $e^{\pm}$ 
$p_T$ distribution. The result is consistent with a full detector 
simulation.

Accurate momentum 
reconstruction in the tracking system requires the precise determination 
of the beam position in the plane orthogonal to the beam line. 
This was measured and monitored using straight tracks from special runs 
with the magnetic field off throughout the data taking period.

%%%%%%%%%%%%%%%%%%%%%%%%%%%%%%%%%%%%%%%%%%%%%%%%%%%%% Fig_1
\begin{figure}[thb]
\includegraphics[width=1.0\linewidth]{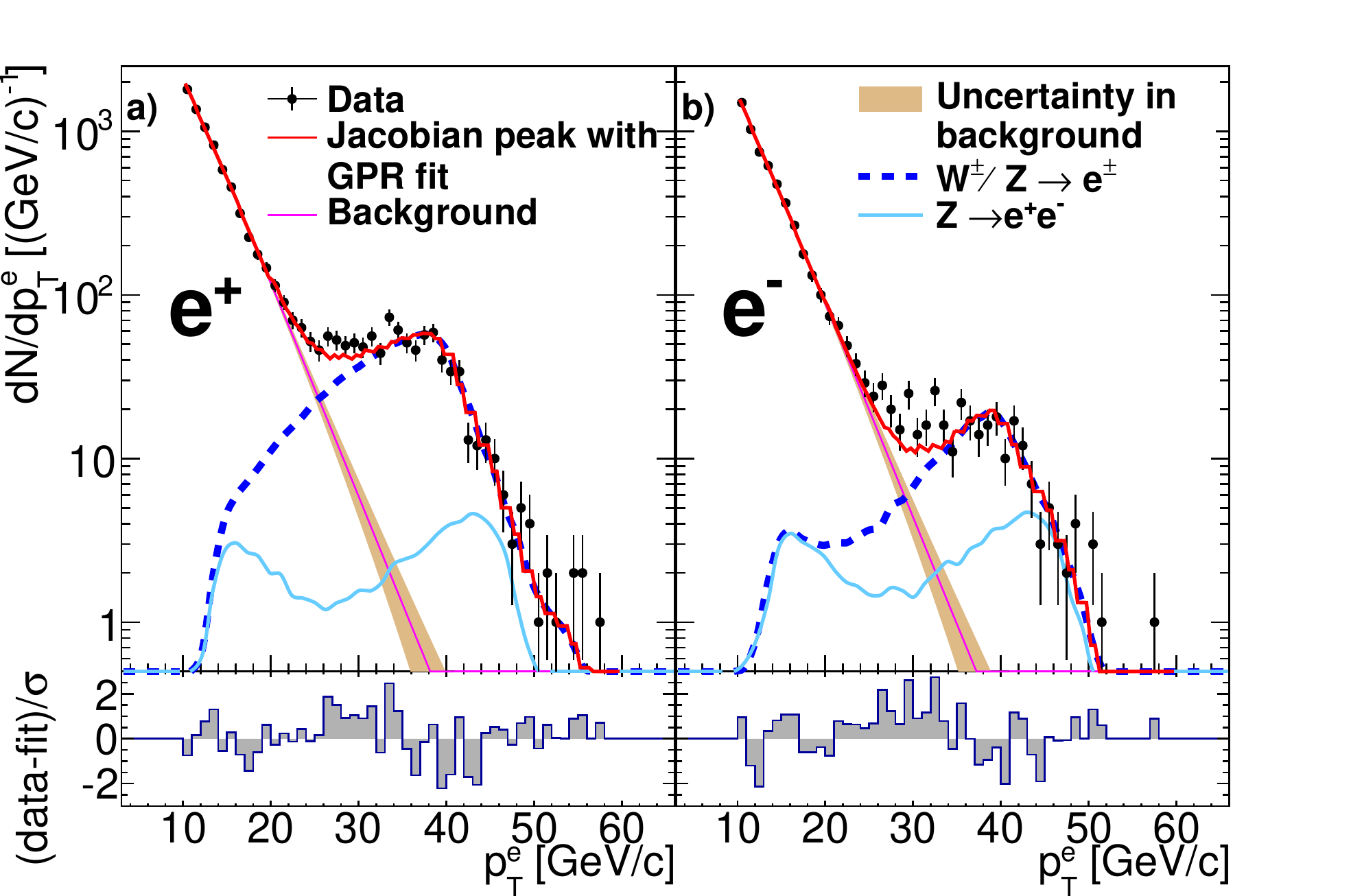}  
 \caption{\label{fig:ptSpectra} (color online).
(Upper panels) Spectra for (a) $e^+$ and (b) $e^-$ using the EMCal for 
momentum determination from $p$$+$$p$ collisions at $\sqrt{s}=510$ GeV 
from 2013.  From top to bottom, the curves are:
solid [red] is the sum between background and signal; 
shaded band [orange] is the background estimation with the 
uncertainty from the GPR method;
dashed [dark blue] is the $W^\pm/Z \rightarrow e^\pm$ signal obtained 
from simulation normalized to the data; 
and solid [light blue] is the contribution from $Z \rightarrow e^+ e^-$.
(Lower panels) Point-by-point comparison of the 
data and the fit result: (data$_i$-fit)/$\sigma_i$, where $\sigma_i$ is 
statistical uncertainty of the $i$-th data point.
 }
 \end{figure}

An isolation cut was very efficient at suppressing background events with 
a high degree of activity around a candidate electron (as would happen for 
jet events). The cut parameter $r_{\rm iso}$ was defined as 
$r_{\rm iso}=\left(\Sigma E_i\right)/E_e$, where $E_i$ is the $i$-th EMCal 
cluster energy and track $p_T$ around the electron candidate in a cone 
with a radius in $\eta$ and $\phi$ of 0.4, and $E_e$ is the energy of 
electron candidate. A candidate was kept for the analysis if $r_{\rm 
iso}<0.1$. 

Figure~\ref{fig:ptSpectra} shows the resulting yield of electron and 
positron candidates for the 2013 data set.  A Jacobian peak around 
$p_{T}^{e}=$40 GeV/$c$ corresponds to $e^{\pm}$ from $W$ and $Z$ boson 
decays. The isolation cut removed about 90\% of the background (as was 
evaluated from the background dominated region between 10 and 20 GeV/$c$) 
and left more than 90\% of the signal in the Jacobian peak region 
untouched (as evaluated from simulations explained below). Similar results 
were obtained for the 2011 and 2012 data sets. Above 30 GeV/$c$ the 
remaining candidate events after the isolation cut are dominated by $W$ 
and $Z$ decay to electrons/positrons, and by background events below 25 
GeV/$c$. This background consists mainly of high momentum 
electrons/positrons from conversion of $\pi^0/\eta$ decay photons, charged 
pions/kaons, $b$,$c\rightarrow e$ decays and accidental matching between 
high energy EMCal clusters and tracks in the DC. The $Z$-boson 
contribution in the signal region above 30 GeV/$c$ was estimated to be 7\% 
(25\%) for the positrons (electrons) after all analysis cuts were applied,
as determined from simulations. 
The asymmetry of the $Z$ has been estimated theoretically using the DSSV08 
PDF sets and measured by the STAR collaboration~\cite{Adamczyk:2014xyw} to 
be $-0.07{\pm}0.14$.

%%%%%%%%%%%%%%%%%%%%%%%%%%%%%%%%%%%%%%%%%%%%%%%%%%%%%%%%%%%%%%%%%%%%%%%%%%%%%%%% Background

%================================================ Table_I
\begin{table*}[htb]
\caption{\label{tab:bkg}
Number of events recorded for $e^+$ and $e^-$ for $30<p_T<50$~GeV/$c$ 
and the background contributions, dilution factors, and two-beam 
polarizations for each analyzed data set.
}
 \begin{ruledtabular} \begin{tabular}{ccccccc}
Lepton & Year & Counts & Background & Dilution   & \multicolumn{2}{c}{Polarization} \\
 &   &   &   Contribution &  Factor & B & Y \\
\hline
$e^+$&2011&70&2.3 $\pm$ 2.3 (stat) $\pm$ 0.6 (syst)&0.97 $\pm$ 0.03 (stat) $\pm$ 0.01 (syst)&0.51 $\pm$ 0.02&0.50 $\pm$ 0.02\\
&2012&105&2.5 $\pm$ 2.5 (stat) $^{+4.7}_{-2.5}$ (syst)&0.98 $\pm$ 0.02 (stat) $^{+0.02}_{-0.04}$ (syst)&0.55 $\pm$ 0.02&0.57 $\pm$ 0.02\\
&2013&669&18.6 $\pm$ 7.3 (stat) $\pm$ 14.9 (syst)&0.97 $\pm$ 0.01 (stat) $\pm$ 0.02 (syst)&0.55 $\pm$ 0.02&0.55 $\pm$ 0.02\\
\\
$e^-$&2011&27&1.7 $\pm$ 1.6 (stat) $\pm$ 0.7 (syst)&0.94 $\pm$ 0.06 (stat) $\pm$ 0.02 (syst)&0.51 $\pm$ 0.02&0.50 $\pm$ 0.02\\
&2012&47&5.5 $\pm$ 4.7 (stat) $\pm$ 2.2 (syst)&0.88 $\pm$ 0.10 (stat) $\pm$ 0.05 (syst)&0.55 $\pm$ 0.02&0.57 $\pm$ 0.02\\
&2013&233&13.9 $\pm$ 5.6 (stat) $^{+20.0}_{-13.9}$ (syst)&0.94 $\pm$ 0.02 (stat) $^{+0.06}_{-0.09}$ (syst)&0.55 $\pm$ 0.02&0.55 $\pm$ 0.02
\\
\end{tabular} \end{ruledtabular}
\end{table*}

Experimentally, the longitudinal single-spin asymmetry is defined as: 
\begin{equation}
A_L = \frac{1}{P} \frac{N^+ - R N^-}{N^+ + R N^-},
\label{eq:ssa}
\end{equation}
where $P$ is the beam polarization, $N^+$ ($N^-$) is the number of events 
in the signal region for the positive (negative) beam helicity and $R$ is 
the luminosity ratio (relative luminosity) between positive and negative 
helicity bunches measured using the minimum bias trigger defined by a 
coincidence of the two beam-beam counters. The relative luminosity between 
different helicity combinations did not differ from unity by more than 
2\%. 
The asymmetry calculation was performed for events in the $p_T$ range from 30 to 50 GeV/$c$, which defined the signal region in this analysis. This range was selected to optimize the signal to background. Less than 1\% of the signal is expected above 50 GeV/$c$.
Asymmetries obtained in this fashion must be corrected for background 
events, which are parity-conserving, in the signal region. This dilution 
factor can be defined as $(A-B)/A$, where $A$ ($B$) is the number of all 
(background) events in the signal region $30<p_T<50$ GeV/$c$. The final 
asymmetry values can be obtained by dividing the result by the dilution 
factor.

The background in the signal region was estimated using the Gaussian 
process regression (GPR) technique~\cite{mackay, rasmus,lauritz,barber} to 
extrapolate the background shape from the background-dominated region to 
the signal-dominated region. The major advantage of this method is that it 
does not require an a priori known functional form to test against data. 
At its core, this method allows for the determination of the shape of a 
set of data points with statistical uncertainties using only the data 
themselves. Furthermore, the predictions made using this method have a 
mathematically well-defined Gaussian uncertainty.

Through our use of the radial basis function (RBF) 
kernel~\cite{mackay,rasmus}, we assume a smooth (infinitely 
differentiable) shape for the background. The background shape was 
constrained from data points in the $p_T$ ranges 10--22 GeV/$c$ and 60--65 
GeV/$c$, where the signal contribution is expected to be negligible. 
Although bins in the range 60--65 GeV/$c$ don't contain any events, they 
still improve the precision of the background evaluation. These empty bins 
were assigned a statistical uncertainty of 1 count. The background in the 
signal region is assumed to vary on $p_{T}$ scales equal or larger than 
those in the background dominated regions.

The RBF kernel contains a characteristic length parameter that is an 
indicator of how far away from data the background extrapolations can be 
made.  For obtained characteristic lengths larger than 30 GeV/$c$, we 
concluded that our background estimation (based on data between 10 to 22 
GeV/$c$ and 60 to 65 GeV/$c$) in the signal region (30 to 50 GeV/$c$) has 
an appropriate statistical uncertainty.

Table~\ref{tab:bkg} summarizes the background contributions with statistical uncertainties obtained using the GPR approach along with the counts in the signal region for each data set. The GPR analysis was performed for different $p_T$ ranges for the background estimation and including/excluding the constraint between 60 and 65 GeV/$c$. The results were within the statistical uncertainty of the full GPR analysis so no additional systematic was added.

In Fig.~\ref{fig:ptSpectra}, the background and signal shapes were used to 
describe the data points. The only fit parameter was the normalization for 
the signal shape. The signal shape was obtained from a {\sc pythia} 
simulation~\cite{Sjostrand:2006za} of $W^\pm$ and $Z$-boson decays to 
electrons/positrons, followed by a full {\sc 
geant}3-based~\cite{Adler:2003zv} detector response simulation. The 
simulated events were analyzed using the analysis package used for the 
data. The fit quality of the data-driven background shape plus the 
simulated signal shape is reasonable for both $e^-$ and $e^+$ spectra.

As a cross check of the background determination, a fit to the data using 
a phenomenologically-motivated modified power law function as the 
background shape was also performed 
$\left(f\left(p_T\right)=1/p_T^{\alpha+\beta \ln{p_T}}\right)$. 
The discrepancy between the central value results from two methods was assigned as a systematic uncertainty for the background determination.
Another source of uncertainty may come from the possible 
systematic discrepancy between the data points and the fit result in some 
$p_T$ regions (e.g. data excess over the fit in the vicinity of $p_T=30$ 
GeV/$c$ in Fig.~\ref{fig:ptSpectra}). Following a conservative approach 
for uncertainty evaluation, the sum of the signed differences between data 
points and the fit results within the signal region was assigned as an 
additional systematic uncertainty.
The final systematic uncertainty was obtained by adding in quadrature the systematic uncertainty from the two sources discussed above. Using the GPR-estimated background contamination in the signal region, the dilution factor for each data set was calculated and is presented in Table~\ref{tab:bkg}.

%%%%%%%%%%%%%%%%%%%%%%%%%%%%%%%%%%%%%%%%%%%%%%%%%%%%%% Asymmetries

The asymmetry calculation was done following two independent methods. 
First the asymmetry was calculated separately for each polarized beam 
using Eq.~\ref{eq:ssa}, with the polarization for the other beam 
averaged to zero. The final result is a weighted average of asymmetries 
from two beams.
A likelihood method was also used in order to deal with the lower 
statistics, particularly in the 2011 and 2012 data sets. 

The two rings at RHIC with counter-propagating beams are designated yellow 
(y,Y) and blue (b,B).  The number of expected counts $\mu_{yb}$ for the 
data sample can be expressed as:

\begin{equation}
\mu_{yb}=R_{yb}N\left(1{+}b{\cdot}{A_L}{P_B}{+}y{\cdot}{A_L}{P_Y}{+}b{\cdot}{y}{\cdot}{A_{LL}}{P_B}{P_Y}\right)
\label{eq:mu}
\end{equation} 

%\nonumber \\
% &b \cdot y \cdot A_{LL} P_B P_Y \right)
   
\noindent where $R_{yb}$ is the relative luminosity between the colliding 
beam helicity configurations, $y$ ($b$) denotes the helicity of the two 
colliding beams and takes the value of $+1$ ($-1$) for positive (negative) 
helicity, the parameter $N$ is an average count, $P_B$ and $P_Y$ are the 
polarizations of the two beams, $A_{LL}$ is the double spin asymmetry. 
The spin asymmetries were calculated by maximizing a likelihood 
function defined using Poisson statistics as:

\begin{equation}
\mathcal{L}=\prod_{y=\pm 1,b=\pm 1} \mathcal{P}\left( \mu_{yb}, N_{yb}\right),
\label{eq:ML}
\end{equation}

\noindent where $N_{yb}$ is the spin sorted yield. To calculate the 2013 
positive and negative $\eta$ bin asymmetries a generalized form for these 
equations was used.
 
%%%%%%%%%%%%%%%%%%%%%%%%%%%%%%%%%%%%%%%%%%%%%%%%% Fig_2
\begin{figure}[thb]
\includegraphics[width=1.0\linewidth]{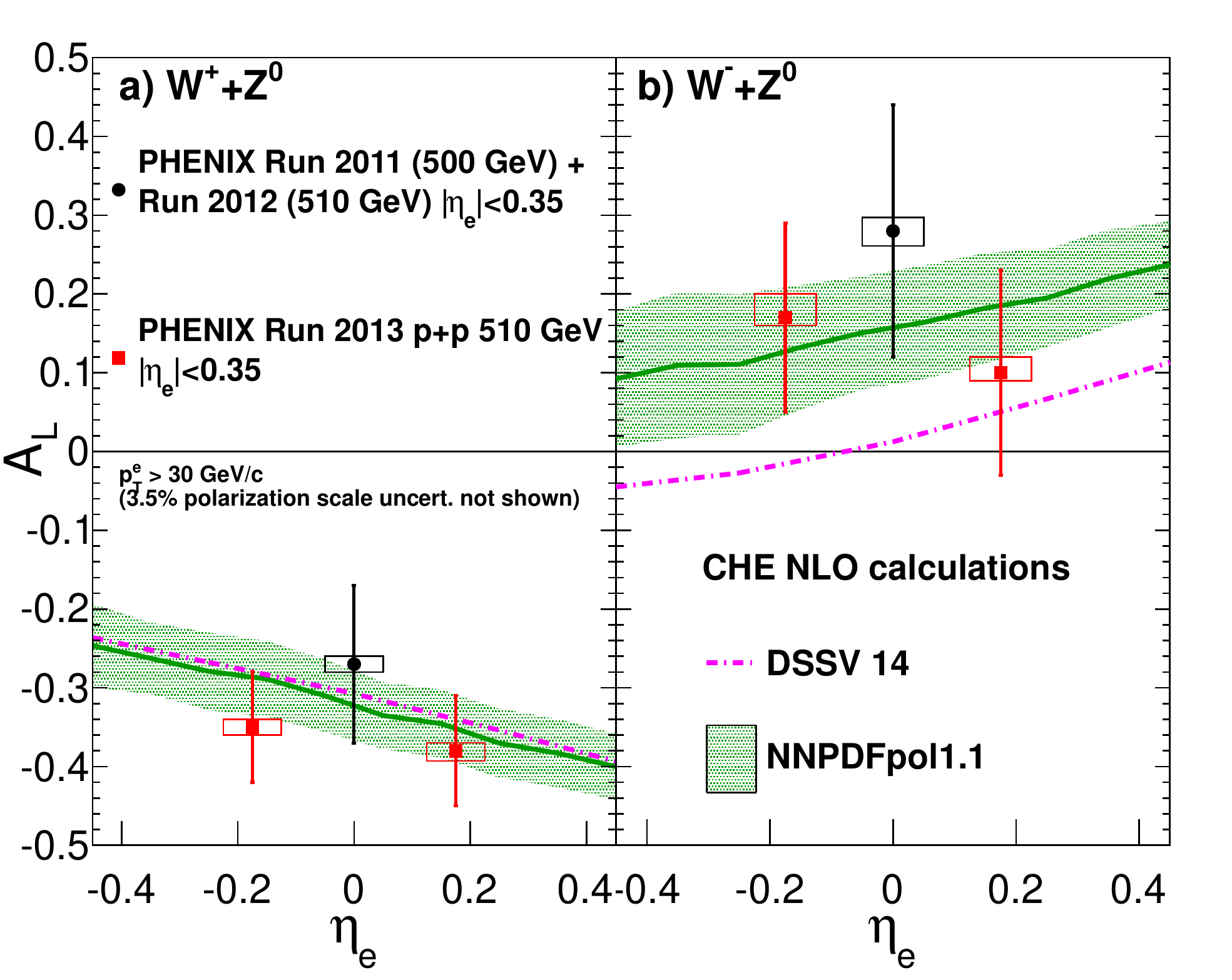}  
 \caption{\label{fig:asym}(color online).
Asymmetry results from the combined 2011 and 2012 data sets for 
$\left|\eta\right|<0.35$ (black circles) and the 2013 data (red squares) 
separated into two equal $\eta$ bins between -0.35 and 0.35. The green 
line and shaded region shows a theoretical calculation using 
CHE~\cite{deFlorian:2010aa} with the NNPDFpol1.1 PDF 
sets~\cite{Nocera:2014gqa}, while the dashed magenta line shows the DSSV14 
calculation~\cite{Ringer:2015oaa}.
 }
\end{figure}

Table~\ref{tab:ssa} summarizes the $A_L$ results.  Both of the asymmetry 
calculation methods employed gave consistent results for all the data 
sets.  
The systematic uncertainties were obtained by propagating the systematic uncertainties of the dilution factors to the final asymmetry values. A scale uncertainty of 3.5\% from the RHIC beam polarization measurements is not included in Table~\ref{tab:bkg}.
The asymmetry in the background region was also measured and for 
all cases the asymmetry was consistent with zero, within uncertainties.

%================================================ Table_II
 \begin{table}[tbh]
 \caption{\label{tab:ssa}
Longitudinal single-spin asymmetries, $A_L$, for the 2011 and 2012 data 
sets (combined) spanning the entire $\eta$ range of PHENIX 
($\left|\eta\right|<0.35$), for the 2013 data set separated 
into two $\eta$ bins, and for the combined 2011-2013 data sets.
 }
 \begin{ruledtabular} \begin{tabular}{cccc}
Lepton & Data Set & $\left<\eta\right>$ & $A_L$ \\\hline
$e^+$ &   2011+2012  & 0 &-0.27  $\pm$  0.10   (stat)  $\pm$  0.01  (syst) \\
&  2013 $\eta>0$ & 0.17 & -0.38  $\pm$  0.07  (stat) $\pm 0.01$  (syst) \\
&  2013 $\eta<0$ & -0.17 & -0.35  $\pm$  0.07  (stat) $\pm 0.01$  (syst) \\ 
&  2011--2013 all & 0 &-0.35  $\pm$  0.04  (stat) $\pm 0.01$  (syst) \\
\\
$e^-$ & 2011+2012  & 0 & 0.28  $\pm$  0.16   (stat)  $\pm$  0.02  (syst)\\
&  2013 $\eta>0$ &0.17 & 0.10  $\pm$  0.13  (stat) $^{+0.02}_{-0.01}$  (syst) \\
&	 2013 $\eta<0$ & -0.17 & 0.17  $\pm$  0.12  (stat) $^{+0.03}_{-0.01}$  (syst) \\  
&  2011--2013 all & 0 & 0.17  $\pm$  0.08  (stat) $\pm 0.02$  (syst) \\
\end{tabular} \end{ruledtabular}
 \end{table}

%2013 $\eta>0$ &  -0.383  $\pm$  0.071  (stat) $^{+0.009}_{-0.014}$  (syst) \\
%2013 $\eta<0$ &  -0.354  $\pm$  0.072  (stat) $^{+0.008}_{-0.014}$  (syst) \\\hline
%	 2013 $\eta>0$ &  0.11  $\pm$  0.13  (stat) $^{+0.020}_{-0.009}$  (syst) \\
%	 2013 $\eta<0$ &  0.17  $\pm$  0.12  (stat) $^{+0.028}_{-0.007}$  (syst) \\  

These results are shown in Fig.~\ref{fig:asym} with two theoretical 
calculations: (CHE)~\cite{deFlorian:2010aa} for the 
NNPDFpol1.1~\cite{Nocera:2014gqa} and a recent 
calculation~\cite{Ringer:2015oaa} using the DSSV 14 PDF 
sets~\cite{deFlorian:2014yva}. While the DSSV 14 curve was obtained from a 
global fit of DIS and SIDIS data (including recent COMPASS 
results~\cite{Alekseev:2010hc,Alekseev:2010ub}), the NNPDFpol1.1 
uncertainty band contains the 2012 STAR~\cite{Adamczyk:2014xyw} result for 
flavor separation in addition to DIS data. The theoretical asymmetry 
calculations agree with the data within 1.5 $\sigma$ uncertainty of the 
data points. These results will be used to further constrain the quark and 
anti-quark polarized parton distributions functions at an intermediate 
Bjorken $x$ value of roughly $M_W/\sqrt{s}=0.16$.

%%%%%%%%%%%%%%%%%%%%%%%%%%%%%%%%%%%%%%%%%%%%%%% Fig_3
\begin{figure}[thb]
\includegraphics[width=1.0\linewidth]{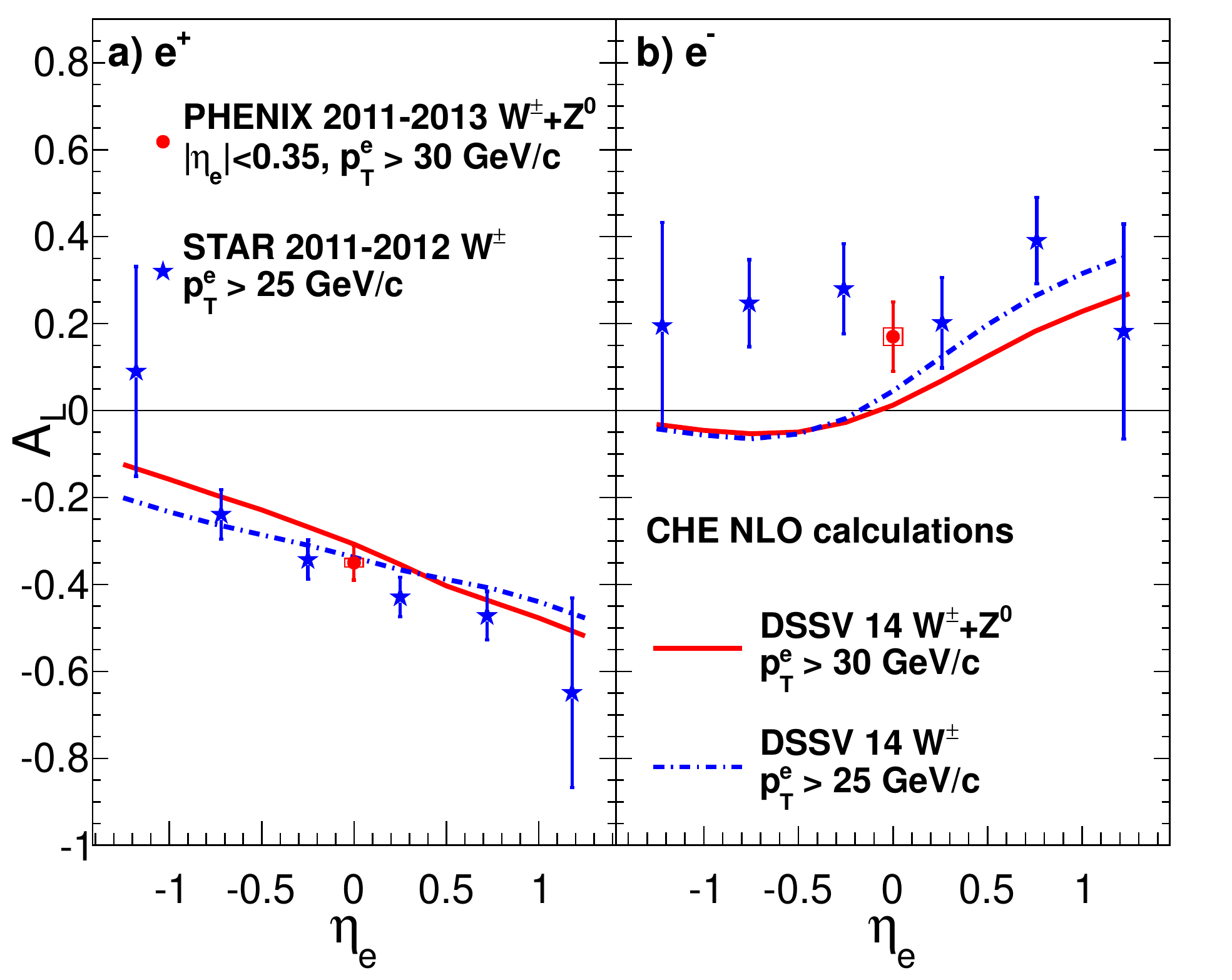}  
 \caption{\label{fig:compare}(color online).
Asymmetry results from the combined 2011--2013 data sets from PHENIX 
[red] circles and the STAR 2011--2012~\cite{Adamczyk:2014xyw} W results 
[blue] stars and their respective DSSV 14 theoretical predictions.
 }
 \end{figure}

Figure~\ref{fig:compare} shows the combined asymmetry for all of the 
PHENIX data sets and published data from STAR~\cite{Adamczyk:2014xyw}. The 
two data sets cannot be compared directly, because PHENIX measures the 
asymmetry from $W^{\pm}+Z$ decays, while the STAR result is solely from 
$W^{\pm}$ decays. The comparison can be made through the curves, which 
account for the specifics of each measurement.  Qualitatively, both data 
sets show the same trend with data points below (above) the central value 
of the theoretical prediction for $W^+$ ($W^-$), for $\left| \eta 
\right|<0.5$. The $W^-$ difference corresponds to a larger $\Delta 
\bar{u}$ contribution in the covered $x \sim 0.16$ range, when compared 
with the central value of the DSSV 14 PDF fit calculation.

%%%%%%%%%%%%%%%%%%%%%%%%%%%%%%%%%%%%%%%%%%%%%%%%%%%%%%%%%%%%% Conclusions

In summary, for high $p_T$ $e^-$ and $e^+$ from $W$ and $Z$ boson decays, 
PHENIX measured the single spin asymmetries with more than 27 times higher 
statistics and better polarization compared to 2009~\cite{Adare:2010xa}. 
These new results and the STAR data~\cite{Adamczyk:2014xyw} 
will help constrain the anti-quark PDFs in a global analysis.  Asymmetries 
calculated from global fits based on previous measurements, such as DSSV14 
and NNPDFpol1.1, are consistent with our data. The use of the electroweak 
interaction provides an independent tool to extract quark and anti-quark 
helicity contribution.  The data presented here are complementary 
to previous SIDIS measurements and bring the field one step closer to 
elucidation of the proton-spin puzzle~\cite{Ashman:1987hv}.

%\textbf{*** page break for PRL word count $<$3.5 pages $<$7 columns} 
%\clearpage

%%%%%%%%%%%%%%%%%%%%%%%%%%%%%%%%%%%%%%%%%%%%%%  Acknowledgments 

%\section*{ACKNOWLEDGMENTS}   % Run-14 long form for all journals

We thank the staff of the Collider-Accelerator and Physics
Departments at Brookhaven National Laboratory and the staff of
the other PHENIX participating institutions for their vital
contributions.  We acknowledge support from the
Office of Nuclear Physics in the
Office of Science of the Department of Energy,
the National Science Foundation,
Abilene Christian University Research Council,
Research Foundation of SUNY, and
Dean of the College of Arts and Sciences, Vanderbilt University
(U.S.A),
Ministry of Education, Culture, Sports, Science, and Technology
and the Japan Society for the Promotion of Science (Japan),
Conselho Nacional de Desenvolvimento Cient\'{\i}fico e
Tecnol{\'o}gico and Funda\c c{\~a}o de Amparo {\`a} Pesquisa do
Estado de S{\~a}o Paulo (Brazil),
Natural Science Foundation of China (People's Republic of~China),
Croatian Science Foundation and
Ministry of Science, Education, and Sports (Croatia),
Ministry of Education, Youth and Sports (Czech Republic),
Centre National de la Recherche Scientifique, Commissariat
{\`a} l'{\'E}nergie Atomique, and Institut National de Physique
Nucl{\'e}aire et de Physique des Particules (France),
Bundesministerium f\"ur Bildung und Forschung, Deutscher
Akademischer Austausch Dienst, and Alexander von Humboldt Stiftung (Germany),
National Science Fund, OTKA, K\'aroly R\'obert University College,
and the Ch. Simonyi Fund (Hungary),
Department of Atomic Energy and Department of Science and Technology (India),
Israel Science Foundation (Israel),
Basic Science Research Program through NRF of the Ministry of Education (Korea),
Physics Department, Lahore University of Management Sciences (Pakistan),
Ministry of Education and Science, Russian Academy of Sciences,
Federal Agency of Atomic Energy (Russia),
VR and Wallenberg Foundation (Sweden),
the U.S. Civilian Research and Development Foundation for the
Independent States of the Former Soviet Union,
the Hungarian American Enterprise Scholarship Fund,
and the US-Israel Binational Science Foundation.

%%%%%%%%%%%%%%%%%%%%%%%%%%%  References 

%\bibliography{ppg169x2}   

%merlin.mbs apsrev4-1.bst 2010-07-25 4.21a (PWD, AO, DPC) hacked
%Control: key (0)
%Control: author (0) dotless jnrlst
%Control: editor formatted (1) identically to author
%Control: production of article title (0) allowed
%Control: page (1) range
%Control: year (0) verbatim
%Control: production of eprint (0) enabled
%
 
\end{document}